\documentclass[twocolumn,showpacs,prb,amsmath,amssymb,aps,floatfix]{revtex4-1}

\usepackage{graphicx,subfigure}
\usepackage{dcolumn}
\usepackage{bm}
\usepackage{gensymb}
\usepackage{wasysym}

\begin{document}

\title{Model and Simulations of the Epitaxial Growth of Graphene on Non-Planar 6H-SiC Surfaces}

\author{Fan Ming}
\author{Andrew Zangwill}%
 \email{andrew.zangwill@physics.gatech.edu}
\affiliation{%
School of Physics, Georgia Institute of Technology, Atlanta, GA 30332, USA
}%

\date{\today}

\begin{abstract}
We study step flow growth of epitaxial graphene on 6H-SiC using a one dimensional kinetic Monte Carlo model. The model parameters are effective energy barriers for the nucleation and propagation of graphene at the SiC steps. When the model is applied to graphene growth on vicinal surfaces, a strip width distribution is used to characterize the surface morphology. Additional kinetic processes are included to study graphene growth on SiC nano-facets. Our main result is that the original nano-facet is fractured into several nano-facets during graphene growth. This phenomenon is characterized by the angle at which the fractured nano-facet is oriented with respect to the basal plane. The distribution of this angle across the surface is found to be related to the strip width distribution for vicinal surfaces. As the terrace propagation barrier decreases, the fracture angle distribution changes continously from two-sided Gaussian to one-sided power-law. Using this distribution, it will be possible to extract energy barriers from experiments and interpret the growth morphology quantitatively.
        
\end{abstract}

\pacs{68.35.-p, 68.55.-a, 81.15.Aa}

\maketitle

\section{Introduction}
Epitaxial graphene grown by thermal decomposition of SiC is an attractive enabling technology for integrating graphene into silicon microelectronics\cite{Berger2004, First2010}. However, scalable production of high quality graphene film is still challenging. Progress in this field requires understanding of the growth mechanism. Unfortunately, most experimental results on epitaxial graphene growth are qualitative. The same is true for the interpretation of these results. 

First-principle approaches to this subject are difficult due to the complexity of the graphene/SiC interface \cite{Hass2008,Tromp2009} and an unknown step structure. Previously, the present authors introduced a simple one dimensional kinetic Monte Carlo (KMC) model to study the epitaxial growth of graphene on vicinal surfaces of 6H-SiC \cite{Ming2011}. Based on a study of graphene strip width distributions, the simulation results showed two distinct growth regimes dominated by ``coalescence" processes and ``climb-over" processes, respectively.  

Besides growth on SiC vicinal substrates, it has been shown recently that graphene grows spontaneously on nano-facetted SiC substrates \cite{Emtsev2009,Virojanadara2010,Robinson2010}. In this context, a ``nano-facet" refers to a stable structure composed of several closely-spaced triple bilayer steps on the flat SiC surface. Suprisingly, perhaps, growth on a nano-facetted surface often leads to better quality graphene films on the adjacent terrace \cite{Emtsev2009,Virojanadara2010}. Growth on nano-facets also plays an important role in graphene ribbon growth\cite{Sprinkle2010}. The purpose of this paper is to extend our previous study for vicinal substrates to include growth on nano-facetted substrates.
       
Commercially available SiC substrates exhibt rough and scratched surfaces. An effective method to produce atomically flat vicinal surfaces is high temperature H$_2$ etching. For on-axis 6H-SiC substrates, this method produces vicinal surfaces with triple bilayer steps. This type of microstep formation is related to etching kinetics and energy differences between different basal planes \cite{Hayashi2009}. This contrasts with off-axis 6H-SiC substrates where the H$_2$ etched surface often shows periodic structures of nano-facets. Depending on whether the substrate is tilted toward the $\langle 1\bar{\rm 1}00\rangle$ or $\langle 11\bar{\rm 2}0\rangle$ direction, the nano-facet makes an angle of $\approx 25^\circ$ or $13-14^\circ$ from the basal plane \cite{Nakajima2005,Nie2008,Nakagawa2003}. Besides H$_2$ etching, controled nano-facets can be achieved by direct plasma etching of SiC surfaces \cite{Sprinkle2010}.

Nano-facets can also form spontaneously when SiC is heated close to the graphitization temperature in non-vacuum environments. This type of step bunching is observed for both on- and off-axis SiC substrates. At higher growth pressure, a reduced silicon sublimation rate leads to a higher graphitization temperature. It is likely that at these elevated temperatures, the SiC vicinal surface of triple bilayer steps is further reconstructed into nano-facets to minimize the surface free energy. The formation of graphene begins at these nano-facets where silicon atoms have fewer bonds \cite{Norimatsu2010,Robinson2010}. Given the previous discussion, SiC step bunching to produce nano-facets may occur before graphene formation. This paper only considers the formation of graphene on pre-existing nano-facets regardless of their origin. We also neglect the difference between $(1\bar{1}0n)$ and $(11\bar{2}n)$ nano-facets. 

The structure of this paper is as follows. Section II is a brief review of our previous results for graphene growth on vicinal surfaces. In section III, we introduce our new model for graphene growth on a SiC surface with nano-facets. Section IV focuses on the distribution of fracture angles which can be compared directly with experiments to extract effective energy barriers.

\section{Strip Width Distribution}
\begin{figure}
\includegraphics[scale=0.38]{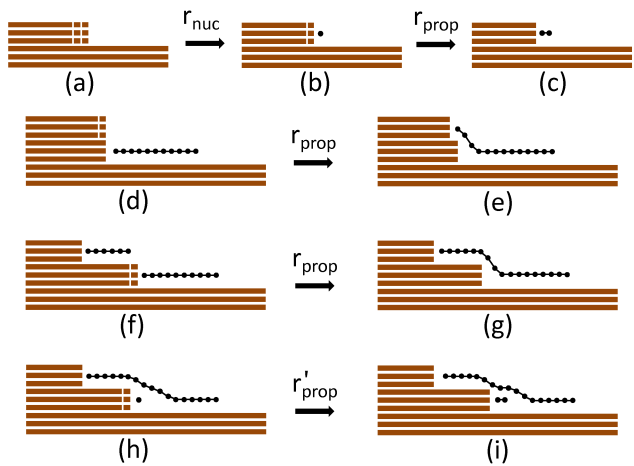}
\caption{Graphene growth kinetic processes on a vicinal surface.}
\label{model}
\end{figure}

Our original model applies to graphene growth on a vicinal surface of 6H-SiC composed of triple bilayer steps. From mass conservation, one graphene unit is obtained by decomposing one SiC step. Step flow growth from SiC triple bilayer steps produces strips of graphene parallel to the SiC step edges. This allows us to model the growth as one-dimensional. The detail of the atomic structure of the graphene/SiC interface does not play a role here, so we do not consider it in the model. Two model energy barriers are necessary: one is the barrier to graphene nucleation at a step edge (Fig.~\ref{model}(a)-(b)). The other is the barrier to graphene strip propagation. Nucleation occurs at a rate $r_{\rm nuc} = \nu_0 \exp(-E_{\rm nuc}/kT)$, where $\nu_0 \approx 10^{12}$ s$^{-1}$ is an attempt frequency and $T$ is the substrate temperature. Propagation occurs (Fig.~\ref{model}(b)-(c)) at a rate $r_{\rm prop} = \nu_0 \exp(-E_{\rm prop}/kT)$. The two barriers are ``effective" energy parameters which account for Si atoms sublimation, C atoms re-crystallization and subsequent graphene growth along the step edge \cite{Tanaka2010,Ohta2010}. 

After a series of propagation steps, two outcomes are possible for each graphene strip. One is that a growing graphene strip runs into another SiC step. We allow the growth to continue onto the adjacent upper terrace. This is called a ``climb-over" process (Fig.~\ref{model}(d)-(e)). The other outcome is that a growing graphene strip meets another graphene strip on the upper terrace. In this case, the two graphene strips connect to each other to form one continuous strip. We call this a ``coalescence" process (Fig.~\ref{model}(f)-(g)). Both climb-over and coalescence produce a step which is covered by a continuous graphene layer. A second graphene layer can nucleate at such a step under the first layer. We allow the subsequent growth of the second layer graphene to continue in the same way as the first layer.         

\begin{figure}
\includegraphics[scale=0.9]{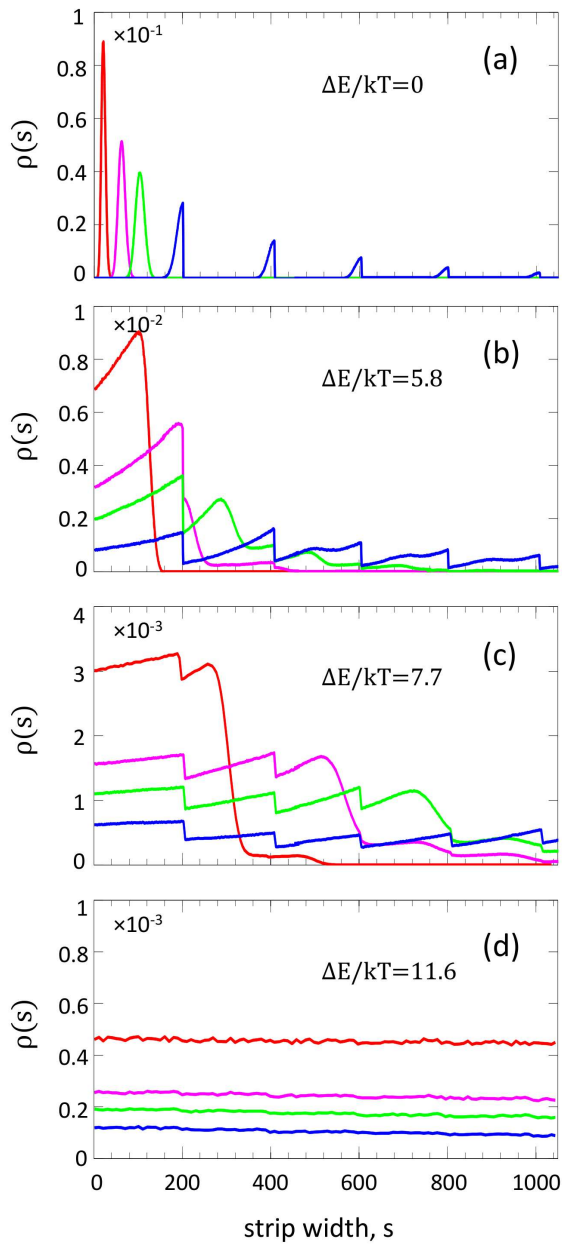}
\caption{Graphene strip width distribution $\rho(s)$ for different $\Delta E$ and total coverages with $\phi=0.9^\circ$. Different color lines correspond to $\Theta=0.1$ (red), $0.3$ (magenta), $0.5$ (green) and $1.0$ (blue), respectively. The terrace width is $W=200$.}
\label{distr}
\end{figure}

Our major finding from this model is the cross-over of the graphene strip width distribution shown in Fig.~\ref{distr}. This cross-over is the result of competition between the climb-over and coalescence processes. We define $\Delta E = E_{\rm nuc} - E_{\rm prop}$ and total coverage $\Theta = \sum_{i} {i\Theta_i}$, where $\Theta_i$ is the graphene coverage of layer i. When the nucleation barrier is small compared to the propagation barrier, i.e. $\Delta E \le 0$, all SiC steps decompose almost simultaneously at the beginning of growth, and the strip width distribution is Poisson (Fig.~\ref{distr}(a)). In this case, coalescence processes occur in a short time interval right before the substrate is fully covered by graphene. Increasing the nucleation barrier relative to the propagation barrier reduces the number of graphene strips on the surface and makes ``climb-over" processes more dominant than coalescences. This decreases the peak magnitude in the strip width distribution and shifts the peak position to the right in Fig.~\ref{distr}(b) and (c). Eventually, when $\Delta E$ is big enough, the step edges become indistinguishable as the graphene strip grows. The strip width distribution is then uniform(Fig.~\ref{distr}(d)).

\section{Modeling Growth on A Nano-facet}
\begin{figure}
\includegraphics[scale=0.45]{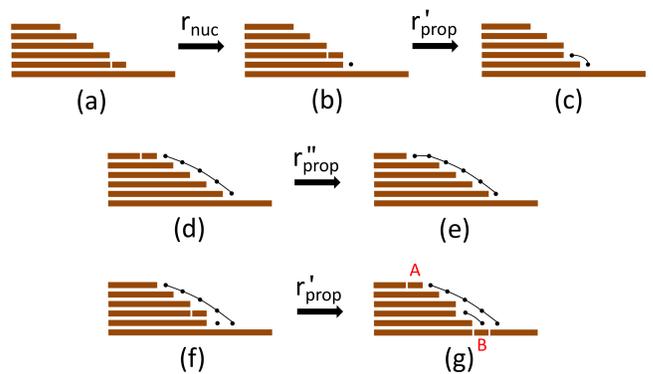}
\caption{Graphene growth kinetics processes on a nano-facet. The blocks specified by letter A and B are used to calculate the fracture angle $\theta$. }
\label{facet}
\end{figure}

In our modeling, a nano-facet is a group of triple bilayer SiC steps with a fixed spacing between adjacent steps (Fig.~\ref{facet}). Hence, the graphene growth proceeds also in step flow mode which can be modeled as one-dimensional. The nano-facet can have different angles with respect to the basal plane depending on both the orientation of the substrate and the graphene growth conditions. Adjusting the spacing width in the model allows different initial nano-facet angles. We assume that growth starts at the bottom of the nano-facet, converting one step into one graphene unit. The nano-facet nucleation process occurs at a rate $r_{\rm nuc} = \nu_0 \exp(-E_{\rm nuc}/kT)$ (Fig.~\ref{facet}(a)-(b)). The nucleated graphene unit propagates upward along the nano-facet at a rate $r_{\rm prop} = \nu_0 \exp(-E_{\rm prop}/kT)$ (Fig.~\ref{facet}(b)-(c)). We keep the value of $E_{\rm prop} = 0$ because incomplete graphene coverage on the nano-facet is rarely observed experimentally. In other words, growth on the nano-facet is very fast. 

The nano-facet nucleation and propagation processes defined here are very similar to those for vicinal surfaces. As soon as a nano-facet propagation event occurs, a second graphene layer can nucleate immediately under the first layer (Fig.~\ref{facet}(f)) and the growth of the second layer can continue in the same way as the first layer (Fig.~\ref{facet}(f)-(g)). This contrasts with our previous model for vicinal surface growth where a second layer of graphene can only grow at a step covered by a continous graphene strip (see Fig.~\ref{model}(h)). When the propagation on the facet reaches the top junction between the nano-facet and the SiC(0001) basal plane, the graphene growth is allowed to continue on the (0001) basal plane but with a slower propagation rate $r'_{\rm prop} = \nu_0 \exp(-E'_{\rm prop}/kT)$ (Fig.~\ref{facet}(d)-(e)). This is due to the experimental observation that several layers of graphene often grow on a nano-facet before the graphene growth propagates on the (0001) basal plane \cite{Robinson2010}. We focus on the early growth stage where the graphene layers of each macrostep grow independently, and no coalescences or climb-overs are allowed.

\begin{figure}
\includegraphics[scale=0.38]{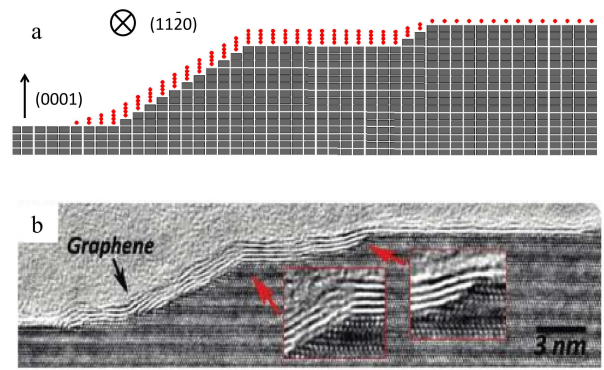}
\caption{(a) KMC simulation snapshot. Total surface coverage $\Theta = 0.8$, $T=1800$ K, $E_{\rm nuc}/kT=7.7$, $E'_{\rm prop}/kT$ is 3.9 and 7.1 for the top graphene layer and the interface graphene layers, respectively. (b) Transmission electron microscope image, adapted from Robinson et al. \cite{Robinson2010}. }
\label{TEM}
\end{figure}

Fig.~\ref{TEM} shows that a particular choice of KMC model parameters produces a simulated morphology very similar to a transmission electron microscope image of graphene growth on a non-planar SiC(0001) surface. The TEM image was taken of a sample which was prepared at 1325 $^\circ$C for $90$ min, with step heights of $5-15$ nm. Both the TEM and the simulation image show a sharp ending for all graphene layers at the bottom of the nano-facet, where the graphene growth starts. Fig.~\ref{TEM} also shows that, from the left to the right side, graphene layers grow continuously over the junction between the high index nano-facet and SiC(0001). The top graphene layer is the longest, while the other layers have a similar length. In what follows, we call all the graphene layers other than the top one ``interface graphene layers".  

\begin{figure}
\includegraphics[scale=0.38]{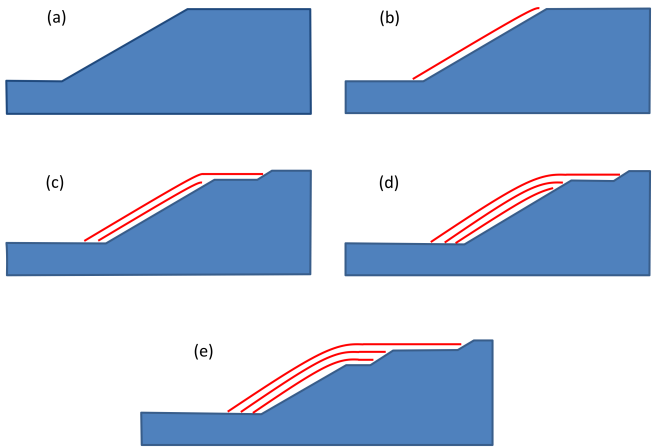}
\caption{Formation processes for the surface morphology in Fig.~\ref{TEM}. }
\label{Formation}
\end{figure}

Fig.~\ref{Formation} illustrates the formation process for the surface morphology in Fig.~\ref{TEM}. The original nano-facet is shown in Fig.~\ref{Formation}(a). The first graphene layer nucleates at the bottom junction of the nano-facet, and propagates quickly up the slope until it reaches the top basal plane. This is shown in Fig.~\ref{Formation}(b). Fig.~\ref{Formation}(c) shows that the first graphene layer continues to grow onto the terrace and a small triple bilayer nano-facet is fractured from the original nano-facet (This process was previously shown in Fig.~\ref{facet}(d)-(e)). However, the terrace propagation rate is slower than the nano-facet propagation rate. Therefore, Fig.~\ref{Formation}(c) also shows that the first graphene layer cannot grow very long on the terrace before the second layer is nucleated under the first layer and quickly covers the lower nano-facet. 

The terrace propagation rate for the second graphene layer is slower than the first layer due to the increasing difficulty for Si atoms to escape from SiC steps covered by graphene. This leads to a possible scenario, shown in Fig.~\ref{Formation}(d), that before the second graphene layer grows on the terrace, a third graphene layer is nucleated and catches up to the second layer growth at the top of the lower nano-facet. Later, when the second graphene layer does grow onto the terrace like the first layer, another triple bilayer step is decomposed from the lower nano-facet. This process is the same for the third layer. However, assuming that the second and third graphene layers have a similar terrace propagation rate, Fig.~\ref{Formation}(e) shows that a nano-facet of one unit cell high is fractured from the original nano-facet by growing the second and third graphene layers together on the terrace.  

Two small nano-facets are now fractured from the origianl nano-facet, with heights corresponding to one and two triple bilayer steps, respectively. This is the same type of surface morphology seen in Fig.~\ref{TEM}(b). We note that, in our model, graphene grows by decomposing SiC triple bilayer steps, so the graphene layer always has one side attached to a nano-facet. To emphasize this, in Fig.~\ref{Formation}, we draw the graphene layers as slightly curved at the top of a nano-facet. 

\section{Fracture Angle Distribution}
To quantify the surface morphology, we study the fracture angle $\theta$ made by the graphene layers on the nano-facet with respect to the basal plane. In the simulations, $\theta$ is quantified as follows. If the nano-facet height is $h$, we count the number of units from the edge of the top terrace (marked as letter A in Fig.~\ref{facet}(g)) to the edge of the bottom terrace (letter B in Fig.~\ref{facet}(g)), and define the length as $L$ (e.g. $L=5$ in Fig.~\ref{facet}(g)). Then $\theta \equiv \arctan{((h+1)/L)}$. The initial $\theta$ in the simulations is assumed to be $\theta_0 = 30^\circ$, which is close to the nano-facet angle observed experimentally. As soon as the top graphene layer growth propagates onto the terrace, $\theta$ starts to decrease as a function of the growth time. Hence, we can use $\theta$ to characterize the nano-facet fracturing process due to graphene growth. $\theta$ can be measured directly by experiments using the definition above. Compared to experimental results, the KMC simulation can be used to provide some quantitative information about the nucleation and propagation barriers. 

The initial surface for our statistical study of growth-induced fracturing is a periodic sequence of basal planes and nano-facets. We fix the height of the nano-facets to be 22.5 nm, which corresponds to $h = 30$ SiC triple bilayers. We use a long terrace width and a short total growth time so that no graphene strip coalescence occurs. For statistical purposes, the surface consists of $6\times 10^4$ alternations of SiC(0001) and nano-facets, with a vicinal angle $\phi = 5.7^\circ$. The growth temperature is fixed at 1800 K. We also fix the nucleation barrier $E_{\rm nuc}/kT=7.7$ and treat the terrace propagation barrier $E'_{\rm prop}$ as the only variable.   

\begin{figure}
\includegraphics[scale=0.38]{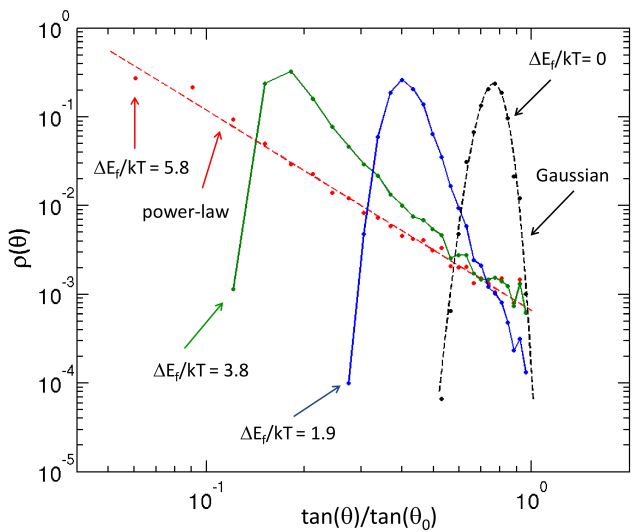}
\caption{Fracture angle distribution for different terrace propagation barriers. The red dashed line is a power-law fit, and the black dashed line is a Gaussian fit. }
\label{angle-dist}
\end{figure}

We now focus on the distribution of fracture angles at a given coverage. For simplicity only, we assume the energy barriers for all graphene layers are the same. Fig.~\ref{angle-dist} shows the distribution $\rho(\theta)$ of the normalized fracture angle $\tan{(\theta)}/\tan{(\theta_0)}$ for different choices of terrace propagation barriers $E'_{\rm prop}$ with a fixed total coverage $\Theta = 0.5$. If we define $\Delta E_f = E_{\rm nuc} - E'_{\rm prop}$, it is clear from Fig.~\ref{angle-dist} that the distribution of fracture angles shows a cross-over behavior as $\Delta E_f$ increases. For $\Delta E_f = 0$, it is difficult for graphene to propagate on the terrace. After all the nano-facets are fully covered by graphene, the growth on the terrace starts almost simultaneously. This leads to a Gaussian distribution of fracture angles in Fig.~\ref{angle-dist}. This growth regime is very similar to graphene growth on vicinal surfaces for $\Delta E=0$, where all the steps are nucleated at the same time and a Poisson strip width distribution is found.

More interestingly, as $\Delta E_f$ increases, Fig.\ref{angle-dist} shows that the distribution of fracture angles becomes more and more one-sided. Eventually, when $\Delta E_f = 5.8$, the distribution follows an intriguing power-law form, in which the probability is zero for $\theta$ values smaller than the minimum $\theta$ given in the figure. We do not fully understand the power-law probability distribution. Nevertheless, the change in fracture angle distribution can still be related to the strip width distribution for vicinal surfaces in Fig.~\ref{distr}. For growth on vicinal surfaces, when the nucleation barrier is high, graphene is nucleated at some steps much earlier than at the others, leading to a shift of peak position to the right side in the strip width distribution (Fig.~\ref{distr}(b)). In other words, longer strips start to dominate the distribution. Similarly, for growth on nano-facets, when $\Delta E_f > 0$, the nano-facets are fractured one after another in a wide time range. In the regime where $\Delta E_f$ is big enough, the distribution is dominated by smaller values of $\theta$ made by longer top graphene layers. This leads to a one-sided distribution of $\theta$.

Given the above discussion, similar to the strip width distribution, the cross-over behavior of the fracture angle distribution can be considered as a competition between the nucleation process at the nano-facet and the propagation process on the terrace. When compared to experimental observations, the fracture angle distribution can be used to determine $\Delta E_f$.

\begin{figure}
\includegraphics[scale=0.38]{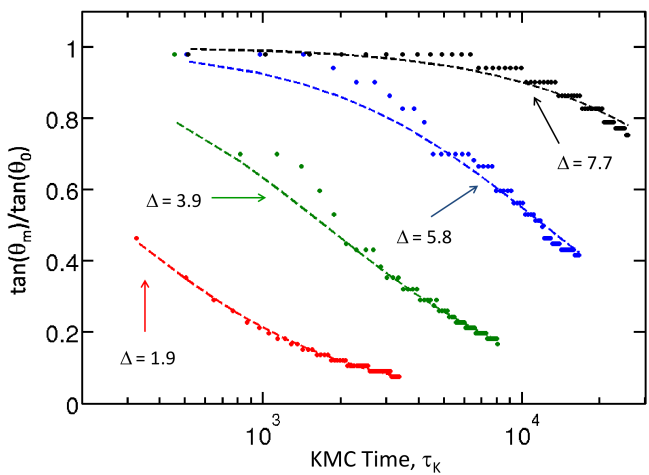}
\caption{$\theta_m$ as a function of KMC time. Dashed curves are the fit results according to Eq.~\ref{angle}. Red, green, blue and black dashed curves give $E''_{\rm prop} /kT = 2.1, 4.0, 6.0$ and $8.0$, respectively. }
\label{fit}
\end{figure}

We define $\theta_m$ as the most probable $\theta$ in the fracture angle distribution. In the following, we give a simple analytic treatment to show how $\theta_m$ evolves as a function of the growth time. If the experimental growth time is $t_E$, the normalized $\theta_m$ can be approximated by
\begin{equation}
\frac{\tan{(\theta_m)}}{\tan{(\theta_0)}} = \frac{h + 1}{h + 1 + \nu_0 t_E \exp(-E''_{\rm prop}/kT)}  .
\label{angle}
\end{equation}      
In this expression, $h$ is the original nano-facet height, and $E''_{\rm prop}$ is an effective terrace propagation barrier. To derive this equation, we do not consider the initial growth time spent on the nano-facet before the terrace growth occurs, which gives a small reduction to $t_E$ in Eq.~\ref{angle}. $E''_{\rm prop}$ accounts for both the top graphene layer terrace propagation barrier $E'_{\rm prop}$ and the nucleation barrier for interface graphene layers. Increasing the number of interface graphene layers will reduce the relative horizontal distance between the top and the bottom terraces at the fractured nano-facet (i.e. the horizontal distance between block A and B in Fig.\ref{facet}(g)). Both the initial growth time spent on the nano-facet and the nucleation of interface graphene layers have the effect of increasing $\theta$. Therefore, it is expected that $E''_{\rm prop} > E'_{\rm prop}$. We treat $E''_{\rm prop}$ as a fitting parameter for our analytic model to see how it is related to $E'_{\rm prop}$. 

Fig.~\ref{fit} shows the simulation result for the normalized $\theta_m$ as a function of the KMC time $\tau_K$ for different choices of terrace propagation barriers. Here we define $\Delta = E'_{\rm prop}/kT$. The KMC time is defined as $\tau_K = t_E r_{prop}$. Dashed lines are fitted curves according to Eq.~\ref{angle}. Despite the simplicity of Eq.~\ref{angle}, the model curves agree quite well with the KMC simulations for smaller $\Delta$ and later growth times. This is likely caused by the adjustment of initial growth time spent on the nano-facet. Smaller $\Delta$ indicates a faster graphene propagation on the terrace, which leads to a relatively shorter growth time spent on the nano-facet before the terrace growth occurs. Similarly, at later growth times when the initial growth time on the nano-facet becomes insignificant, the fitted curve also agrees better. 

As we increase the terrace propagation barrier $E'_{\rm prop}$, the fitting parameter $E''_{\rm prop}$ is also found to increase. In fact, we find an excellent linear relationship between $E'_{\rm prop}$ and $E''_{\rm prop}$: $E''_{\rm prop} = E'_{\rm prop} + 0.03$ eV. The relatively small correction to $E'_{\rm prop}$ suggests that the major contribution of $E''_{\rm prop}$ still comes from $E'_{\rm prop}$ for the top graphene layer. The contribution from the nucleation barrier for interface graphene layers acts as an additional energy barrier $\Delta E_{\rm prop} = 0.03$ eV. Therefore, with Eq.~\ref{angle}, the time evolution of the fracture angle $\theta_m$ measured in an experiment can be used to obtain a good estimate of $E'_{\rm prop}$. Combined with the result for $\Delta E_f$ from the analysis of fracture angle distribution, both $E_{\rm nuc}$ and $E'_{\rm prop}$ can be extracted experimentally. 

Most experiments of graphene growth on SiC nano-facets are done at a temperature between 1600 K and 1800 K. We tested $\Delta E_{\rm prop}$ in this range, but no temperature dependence was found. $\Delta E_{\rm prop}$ also does not seem to be dependent on the original nano-facet height $h$. As we increase the nano-facet height from 22.5 nm to 100 nm, the change of $\Delta E_{\rm prop}$ is less than $10\%$, well within the accuracy of $E''_{\rm prop}$ as obtained from Fig.~\ref{fit}. This suggests that $\Delta E_{\rm prop}$ is only a function of the energy barriers in the model, and very weakly dependent on the growth temperature or the initial nano-facet height in the experimental parameter range.

\section{Conclusion}
We have presented a model for graphene growth on non-planar nano-facetted 6H-SiC substrates based on a previous model of growth on vicinal substrates. The simulation produces a surface morphology very similar to observations. A description of the formation process for this type of surface morphology is also provided. For graphene growth on nano-facetted SiC substrates, a fracture angle can be used to characterize the growth-induced fracture of a nano-facet. The distribution of fracture angles is found to be related to the graphene strip width distribution for growth on vicinal substrates. As the terrace propagation barrier decreases, the fracture angle distribution deviates from a Gaussian and eventually becomes a power-law. Our analytic result for the most probable fracture angle agrees well with the KMC simulations.

\section{Acknowledgement}
We thank Ed Conrad, Phil First and Walt de Heer for useful discussions. Fan Ming was supported by the MRSEC program of the National Science Foundation under Grant No. DMR-0820382.

\bibliography{Graphene-Facet}

\begin{thebibliography}{16}%
\makeatletter
\providecommand \@ifxundefined [1]{%
 \@ifx{#1\undefined}
}%
\providecommand \@ifnum [1]{%
 \ifnum #1\expandafter \@firstoftwo
 \else \expandafter \@secondoftwo
 \fi
}%
\providecommand \@ifx [1]{%
 \ifx #1\expandafter \@firstoftwo
 \else \expandafter \@secondoftwo
 \fi
}%
\providecommand \natexlab [1]{#1}%
\providecommand \enquote  [1]{``#1''}%
\providecommand \bibnamefont  [1]{#1}%
\providecommand \bibfnamefont [1]{#1}%
\providecommand \citenamefont [1]{#1}%
\providecommand \href@noop [0]{\@secondoftwo}%
\providecommand \href [0]{\begingroup \@sanitize@url \@href}%
\providecommand \@href[1]{\@@startlink{#1}\@@href}%
\providecommand \@@href[1]{\endgroup#1\@@endlink}%
\providecommand \@sanitize@url [0]{\catcode `\\12\catcode `\$12\catcode
  `\&12\catcode `\#12\catcode `\^12\catcode `\_12\catcode `\%12\relax}%
\providecommand \@@startlink[1]{}%
\providecommand \@@endlink[0]{}%
\providecommand \url  [0]{\begingroup\@sanitize@url \@url }%
\providecommand \@url [1]{\endgroup\@href {#1}{\urlprefix }}%
\providecommand \urlprefix  [0]{URL }%
\providecommand \Eprint [0]{\href }%
\providecommand \doibase [0]{http://dx.doi.org/}%
\providecommand \selectlanguage [0]{\@gobble}%
\providecommand \bibinfo  [0]{\@secondoftwo}%
\providecommand \bibfield  [0]{\@secondoftwo}%
\providecommand \translation [1]{[#1]}%
\providecommand \BibitemOpen [0]{}%
\providecommand \bibitemStop [0]{}%
\providecommand \bibitemNoStop [0]{.\EOS\space}%
\providecommand \EOS [0]{\spacefactor3000\relax}%
\providecommand \BibitemShut  [1]{\csname bibitem#1\endcsname}%
\let\auto@bib@innerbib\@empty
\bibitem [{\citenamefont {Berger}\ \emph {et~al.}(2004)\citenamefont {Berger},
  \citenamefont {Song}, \citenamefont {Li}, \citenamefont {Li}, \citenamefont
  {Ogbazghi}, \citenamefont {Feng}, \citenamefont {Dai}, \citenamefont
  {Marchenkov}, \citenamefont {Conrad}, \citenamefont {First},\ and\
  \citenamefont {de~Heer}}]{Berger2004}%
  \BibitemOpen
  \bibfield  {author} {\bibinfo {author} {\bibfnamefont {C.}~\bibnamefont
  {Berger}}, \bibinfo {author} {\bibfnamefont {Z.~M.}\ \bibnamefont {Song}},
  \bibinfo {author} {\bibfnamefont {T.~B.}\ \bibnamefont {Li}}, \bibinfo
  {author} {\bibfnamefont {X.~B.}\ \bibnamefont {Li}}, \bibinfo {author}
  {\bibfnamefont {A.~Y.}\ \bibnamefont {Ogbazghi}}, \bibinfo {author}
  {\bibfnamefont {R.}~\bibnamefont {Feng}}, \bibinfo {author} {\bibfnamefont
  {Z.~T.}\ \bibnamefont {Dai}}, \bibinfo {author} {\bibfnamefont {A.~N.}\
  \bibnamefont {Marchenkov}}, \bibinfo {author} {\bibfnamefont {E.~H.}\
  \bibnamefont {Conrad}}, \bibinfo {author} {\bibfnamefont {P.~N.}\
  \bibnamefont {First}}, \ and\ \bibinfo {author} {\bibfnamefont {W.~A.}\
  \bibnamefont {de~Heer}},\ }\href@noop {} {\bibfield  {journal} {\bibinfo
  {journal} {J. Phys. Chem. B}\ }\textbf {\bibinfo {volume} {108}},\ \bibinfo
  {pages} {19912} (\bibinfo {year} {2004})}\BibitemShut {NoStop}%
\bibitem [{\citenamefont {First}\ \emph {et~al.}(2010)\citenamefont {First},
  \citenamefont {de~Heer}, \citenamefont {Seyller}, \citenamefont {Berger},
  \citenamefont {Stroscio},\ and\ \citenamefont {Moon}}]{First2010}%
  \BibitemOpen
  \bibfield  {author} {\bibinfo {author} {\bibfnamefont {P.~N.}\ \bibnamefont
  {First}}, \bibinfo {author} {\bibfnamefont {W.~A.}\ \bibnamefont {de~Heer}},
  \bibinfo {author} {\bibfnamefont {T.}~\bibnamefont {Seyller}}, \bibinfo
  {author} {\bibfnamefont {C.}~\bibnamefont {Berger}}, \bibinfo {author}
  {\bibfnamefont {J.~A.}\ \bibnamefont {Stroscio}}, \ and\ \bibinfo {author}
  {\bibfnamefont {J.~S.}\ \bibnamefont {Moon}},\ }\href@noop {} {\bibfield
  {journal} {\bibinfo  {journal} {MRS. Bull.}\ }\textbf {\bibinfo {volume}
  {35}},\ \bibinfo {pages} {296} (\bibinfo {year} {2010})}\BibitemShut
  {NoStop}%
\bibitem [{\citenamefont {Hass}\ \emph {et~al.}(2008)\citenamefont {Hass},
  \citenamefont {de~Heer},\ and\ \citenamefont {Conrad}}]{Hass2008}%
  \BibitemOpen
  \bibfield  {author} {\bibinfo {author} {\bibfnamefont {J.}~\bibnamefont
  {Hass}}, \bibinfo {author} {\bibfnamefont {W.~A.}\ \bibnamefont {de~Heer}}, \
  and\ \bibinfo {author} {\bibfnamefont {E.~H.}\ \bibnamefont {Conrad}},\
  }\href@noop {} {\bibfield  {journal} {\bibinfo  {journal} {J. Phys.: Cond.
  Matter}\ }\textbf {\bibinfo {volume} {20}},\ \bibinfo {pages} {323202}
  (\bibinfo {year} {2008})}\BibitemShut {NoStop}%
\bibitem [{\citenamefont {Tromp}\ and\ \citenamefont
  {Hannon}(2009)}]{Tromp2009}%
  \BibitemOpen
  \bibfield  {author} {\bibinfo {author} {\bibfnamefont {R.~M.}\ \bibnamefont
  {Tromp}}\ and\ \bibinfo {author} {\bibfnamefont {J.~B.}\ \bibnamefont
  {Hannon}},\ }\href@noop {} {\bibfield  {journal} {\bibinfo  {journal} {Phys.
  Rev. Lett.}\ }\textbf {\bibinfo {volume} {102}},\ \bibinfo {pages} {106104}
  (\bibinfo {year} {2009})}\BibitemShut {NoStop}%
\bibitem [{\citenamefont {Ming}\ and\ \citenamefont
  {Zangwill}(2011)}]{Ming2011}%
  \BibitemOpen
  \bibfield  {author} {\bibinfo {author} {\bibfnamefont {F.}~\bibnamefont
  {Ming}}\ and\ \bibinfo {author} {\bibfnamefont {A.}~\bibnamefont
  {Zangwill}},\ }\href@noop {} {\bibfield  {journal} {\bibinfo  {journal}
  {arXiv:1011.4096}\ } (\bibinfo {year} {2011})}\BibitemShut {NoStop}%
\bibitem [{\citenamefont {Emtsev}\ \emph {et~al.}(2009)\citenamefont {Emtsev},
  \citenamefont {Bostwick}, \citenamefont {Horn}, \citenamefont {Jobst},
  \citenamefont {Kellogg}, \citenamefont {Ley}, \citenamefont {McChesney},
  \citenamefont {Ohta}, \citenamefont {Reshanov}, \citenamefont {Rohrl},
  \citenamefont {Rotenberg}, \citenamefont {Schmid}, \citenamefont {Waldmann},
  \citenamefont {Weber},\ and\ \citenamefont {Seyller}}]{Emtsev2009}%
  \BibitemOpen
  \bibfield  {author} {\bibinfo {author} {\bibfnamefont {K.~V.}\ \bibnamefont
  {Emtsev}}, \bibinfo {author} {\bibfnamefont {A.}~\bibnamefont {Bostwick}},
  \bibinfo {author} {\bibfnamefont {K.}~\bibnamefont {Horn}}, \bibinfo {author}
  {\bibfnamefont {J.}~\bibnamefont {Jobst}}, \bibinfo {author} {\bibfnamefont
  {G.~L.}\ \bibnamefont {Kellogg}}, \bibinfo {author} {\bibfnamefont
  {L.}~\bibnamefont {Ley}}, \bibinfo {author} {\bibfnamefont {J.~L.}\
  \bibnamefont {McChesney}}, \bibinfo {author} {\bibfnamefont {T.}~\bibnamefont
  {Ohta}}, \bibinfo {author} {\bibfnamefont {S.~A.}\ \bibnamefont {Reshanov}},
  \bibinfo {author} {\bibfnamefont {J.}~\bibnamefont {Rohrl}}, \bibinfo
  {author} {\bibfnamefont {E.}~\bibnamefont {Rotenberg}}, \bibinfo {author}
  {\bibfnamefont {A.~K.}\ \bibnamefont {Schmid}}, \bibinfo {author}
  {\bibfnamefont {D.}~\bibnamefont {Waldmann}}, \bibinfo {author}
  {\bibfnamefont {H.~B.}\ \bibnamefont {Weber}}, \ and\ \bibinfo {author}
  {\bibfnamefont {T.}~\bibnamefont {Seyller}},\ }\href@noop {} {\bibfield
  {journal} {\bibinfo  {journal} {Nat. Mater.}\ }\textbf {\bibinfo {volume}
  {8}},\ \bibinfo {pages} {203} (\bibinfo {year} {2009})}\BibitemShut {NoStop}%
\bibitem [{\citenamefont {Virojanadara}\ \emph {et~al.}(2010)\citenamefont
  {Virojanadara}, \citenamefont {Yakimova}, \citenamefont {Zakharov},\ and\
  \citenamefont {Johansson}}]{Virojanadara2010}%
  \BibitemOpen
  \bibfield  {author} {\bibinfo {author} {\bibfnamefont {C.}~\bibnamefont
  {Virojanadara}}, \bibinfo {author} {\bibfnamefont {R.}~\bibnamefont
  {Yakimova}}, \bibinfo {author} {\bibfnamefont {A.~A.}\ \bibnamefont
  {Zakharov}}, \ and\ \bibinfo {author} {\bibfnamefont {L.~I.}\ \bibnamefont
  {Johansson}},\ }\href@noop {} {\bibfield  {journal} {\bibinfo  {journal} {J.
  Phys. D: Appl. Phys.}\ }\textbf {\bibinfo {volume} {43}},\ \bibinfo {pages}
  {374010} (\bibinfo {year} {2010})}\BibitemShut {NoStop}%
\bibitem [{\citenamefont {Robinson}\ \emph {et~al.}(2010)\citenamefont
  {Robinson}, \citenamefont {Weng}, \citenamefont {Trumbull}, \citenamefont
  {Cavalero}, \citenamefont {Wetherington}, \citenamefont {Frantz},
  \citenamefont {LaBella}, \citenamefont {Hughes}, \citenamefont {Fanton},\
  and\ \citenamefont {Snyder}}]{Robinson2010}%
  \BibitemOpen
  \bibfield  {author} {\bibinfo {author} {\bibfnamefont {J.}~\bibnamefont
  {Robinson}}, \bibinfo {author} {\bibfnamefont {X.~J.}\ \bibnamefont {Weng}},
  \bibinfo {author} {\bibfnamefont {K.}~\bibnamefont {Trumbull}}, \bibinfo
  {author} {\bibfnamefont {R.}~\bibnamefont {Cavalero}}, \bibinfo {author}
  {\bibfnamefont {M.}~\bibnamefont {Wetherington}}, \bibinfo {author}
  {\bibfnamefont {E.}~\bibnamefont {Frantz}}, \bibinfo {author} {\bibfnamefont
  {M.}~\bibnamefont {LaBella}}, \bibinfo {author} {\bibfnamefont
  {Z.}~\bibnamefont {Hughes}}, \bibinfo {author} {\bibfnamefont
  {M.}~\bibnamefont {Fanton}}, \ and\ \bibinfo {author} {\bibfnamefont
  {D.}~\bibnamefont {Snyder}},\ }\href@noop {} {\bibfield  {journal} {\bibinfo
  {journal} {Acs Nano}\ }\textbf {\bibinfo {volume} {4}},\ \bibinfo {pages}
  {153} (\bibinfo {year} {2010})}\BibitemShut {NoStop}%
\bibitem [{\citenamefont {Sprinkle}\ \emph {et~al.}(2010)\citenamefont
  {Sprinkle}, \citenamefont {Ruan}, \citenamefont {Hu}, \citenamefont
  {Hankinson}, \citenamefont {Rubio-Roy}, \citenamefont {Zhang}, \citenamefont
  {Wu}, \citenamefont {Berger},\ and\ \citenamefont {de~Heer}}]{Sprinkle2010}%
  \BibitemOpen
  \bibfield  {author} {\bibinfo {author} {\bibfnamefont {M.}~\bibnamefont
  {Sprinkle}}, \bibinfo {author} {\bibfnamefont {M.}~\bibnamefont {Ruan}},
  \bibinfo {author} {\bibfnamefont {Y.}~\bibnamefont {Hu}}, \bibinfo {author}
  {\bibfnamefont {J.}~\bibnamefont {Hankinson}}, \bibinfo {author}
  {\bibfnamefont {M.}~\bibnamefont {Rubio-Roy}}, \bibinfo {author}
  {\bibfnamefont {B.}~\bibnamefont {Zhang}}, \bibinfo {author} {\bibfnamefont
  {X.}~\bibnamefont {Wu}}, \bibinfo {author} {\bibfnamefont {C.}~\bibnamefont
  {Berger}}, \ and\ \bibinfo {author} {\bibfnamefont {W.~A.}\ \bibnamefont
  {de~Heer}},\ }\href@noop {} {\bibfield  {journal} {\bibinfo  {journal} {Nat.
  Nanotech.}\ }\textbf {\bibinfo {volume} {5}},\ \bibinfo {pages} {727}
  (\bibinfo {year} {2010})}\BibitemShut {NoStop}%
\bibitem [{\citenamefont {Hayashi}\ \emph {et~al.}(2009)\citenamefont
  {Hayashi}, \citenamefont {Morita}, \citenamefont {Mizuno}, \citenamefont
  {Tochihara},\ and\ \citenamefont {Tanaka}}]{Hayashi2009}%
  \BibitemOpen
  \bibfield  {author} {\bibinfo {author} {\bibfnamefont {K.}~\bibnamefont
  {Hayashi}}, \bibinfo {author} {\bibfnamefont {K.}~\bibnamefont {Morita}},
  \bibinfo {author} {\bibfnamefont {S.}~\bibnamefont {Mizuno}}, \bibinfo
  {author} {\bibfnamefont {H.}~\bibnamefont {Tochihara}}, \ and\ \bibinfo
  {author} {\bibfnamefont {S.}~\bibnamefont {Tanaka}},\ }\href@noop {}
  {\bibfield  {journal} {\bibinfo  {journal} {Surf. Sci.}\ }\textbf {\bibinfo
  {volume} {603}},\ \bibinfo {pages} {566} (\bibinfo {year}
  {2009})}\BibitemShut {NoStop}%
\bibitem [{\citenamefont {Nakajima}\ \emph {et~al.}(2005)\citenamefont
  {Nakajima}, \citenamefont {Yokoya}, \citenamefont {Furukawa},\ and\
  \citenamefont {Yonezu}}]{Nakajima2005}%
  \BibitemOpen
  \bibfield  {author} {\bibinfo {author} {\bibfnamefont {A.}~\bibnamefont
  {Nakajima}}, \bibinfo {author} {\bibfnamefont {H.}~\bibnamefont {Yokoya}},
  \bibinfo {author} {\bibfnamefont {Y.}~\bibnamefont {Furukawa}}, \ and\
  \bibinfo {author} {\bibfnamefont {H.}~\bibnamefont {Yonezu}},\ }\href@noop {}
  {\bibfield  {journal} {\bibinfo  {journal} {J. Appl. Phys.}\ }\textbf
  {\bibinfo {volume} {97}},\ \bibinfo {pages} {104919} (\bibinfo {year}
  {2005})}\BibitemShut {NoStop}%
\bibitem [{\citenamefont {Nie}\ \emph {et~al.}(2008)\citenamefont {Nie},
  \citenamefont {Lee}, \citenamefont {Feenstra}, \citenamefont {Ke},
  \citenamefont {Devaty}, \citenamefont {Choyke}, \citenamefont {Inoki},
  \citenamefont {Kuan},\ and\ \citenamefont {Gu}}]{Nie2008}%
  \BibitemOpen
  \bibfield  {author} {\bibinfo {author} {\bibfnamefont {S.}~\bibnamefont
  {Nie}}, \bibinfo {author} {\bibfnamefont {C.~D.}\ \bibnamefont {Lee}},
  \bibinfo {author} {\bibfnamefont {R.~M.}\ \bibnamefont {Feenstra}}, \bibinfo
  {author} {\bibfnamefont {Y.}~\bibnamefont {Ke}}, \bibinfo {author}
  {\bibfnamefont {R.~P.}\ \bibnamefont {Devaty}}, \bibinfo {author}
  {\bibfnamefont {W.~J.}\ \bibnamefont {Choyke}}, \bibinfo {author}
  {\bibfnamefont {C.~K.}\ \bibnamefont {Inoki}}, \bibinfo {author}
  {\bibfnamefont {T.~S.}\ \bibnamefont {Kuan}}, \ and\ \bibinfo {author}
  {\bibfnamefont {G.}~\bibnamefont {Gu}},\ }\href@noop {} {\bibfield  {journal}
  {\bibinfo  {journal} {Surf. Sci.}\ }\textbf {\bibinfo {volume} {602}},\
  \bibinfo {pages} {2936} (\bibinfo {year} {2008})}\BibitemShut {NoStop}%
\bibitem [{\citenamefont {Nakagawa}\ \emph {et~al.}(2003)\citenamefont
  {Nakagawa}, \citenamefont {Tanaka},\ and\ \citenamefont
  {Suemune}}]{Nakagawa2003}%
  \BibitemOpen
  \bibfield  {author} {\bibinfo {author} {\bibfnamefont {H.}~\bibnamefont
  {Nakagawa}}, \bibinfo {author} {\bibfnamefont {S.}~\bibnamefont {Tanaka}}, \
  and\ \bibinfo {author} {\bibfnamefont {I.}~\bibnamefont {Suemune}},\
  }\href@noop {} {\bibfield  {journal} {\bibinfo  {journal} {Phys. Rev. Lett.}\
  }\textbf {\bibinfo {volume} {91}},\ \bibinfo {pages} {226107} (\bibinfo
  {year} {2003})}\BibitemShut {NoStop}%
\bibitem [{\citenamefont {Norimatsu}\ and\ \citenamefont
  {Kusunoki}(2010)}]{Norimatsu2010}%
  \BibitemOpen
  \bibfield  {author} {\bibinfo {author} {\bibfnamefont {W.}~\bibnamefont
  {Norimatsu}}\ and\ \bibinfo {author} {\bibfnamefont {M.}~\bibnamefont
  {Kusunoki}},\ }\href@noop {} {\bibfield  {journal} {\bibinfo  {journal}
  {Physica E}\ }\textbf {\bibinfo {volume} {42}},\ \bibinfo {pages} {691}
  (\bibinfo {year} {2010})}\BibitemShut {NoStop}%
\bibitem [{\citenamefont {Tanaka}\ \emph {et~al.}(2010)\citenamefont {Tanaka},
  \citenamefont {Morita},\ and\ \citenamefont {Hibino}}]{Tanaka2010}%
  \BibitemOpen
  \bibfield  {author} {\bibinfo {author} {\bibfnamefont {S.}~\bibnamefont
  {Tanaka}}, \bibinfo {author} {\bibfnamefont {K.}~\bibnamefont {Morita}}, \
  and\ \bibinfo {author} {\bibfnamefont {H.}~\bibnamefont {Hibino}},\
  }\href@noop {} {\bibfield  {journal} {\bibinfo  {journal} {Phys. Rev. B}\
  }\textbf {\bibinfo {volume} {81}},\ \bibinfo {pages} {041406} (\bibinfo
  {year} {2010})}\BibitemShut {NoStop}%
\bibitem [{\citenamefont {Ohta}\ \emph {et~al.}(2010)\citenamefont {Ohta},
  \citenamefont {Bartelt}, \citenamefont {Nie}, \citenamefont {Thurmer},\ and\
  \citenamefont {Kellogg}}]{Ohta2010}%
  \BibitemOpen
  \bibfield  {author} {\bibinfo {author} {\bibfnamefont {T.}~\bibnamefont
  {Ohta}}, \bibinfo {author} {\bibfnamefont {N.~C.}\ \bibnamefont {Bartelt}},
  \bibinfo {author} {\bibfnamefont {S.}~\bibnamefont {Nie}}, \bibinfo {author}
  {\bibfnamefont {K.}~\bibnamefont {Thurmer}}, \ and\ \bibinfo {author}
  {\bibfnamefont {G.~L.}\ \bibnamefont {Kellogg}},\ }\href@noop {} {\bibfield
  {journal} {\bibinfo  {journal} {Phys. Rev. B}\ }\textbf {\bibinfo {volume}
  {81}},\ \bibinfo {pages} {121411} (\bibinfo {year} {2010})}\BibitemShut
  {NoStop}%
\end{thebibliography}%

\end{document}